\documentclass[12pt,a4paper]{article}
\usepackage{graphicx}
\title{The Rare Radiative Annihilation Decays $\bar{B}^0_{s,d} \to J/\psi\gamma$ }

\author{Gongru Lu, Rumin Wang, Y. D. Yang
\thanks{Corresponding author. E-mail address:
yangyd@henannu.edu.cn}\\
{\small {\it Department of Physics, Henan Normal University,
Xinxiang, Henan 453002, P.R. China}} \vspace{2cm}
  }

\begin{document}
\maketitle
\vspace{1.0cm}

\begin{abstract}
 We investigate the physics potential of the annihilation decays
 $\bar{B}^0_{s,d} \to J/\psi$ $\gamma$ in the Standard Model and beyond.
In naive factorization approach, the branching ratios are
estimated to be $\mathcal{B}(\bar{B}^0_s \to
J/\psi\gamma)=1.40\times 10^{-6}$ and $\mathcal{B}(\bar{B}^0_d \to
J/\psi\gamma)=5.29\times
 10^{-8}$. In the framework of QCD factorization, we compute
 the non-factorizable corrections and get
$\mathcal{B}(\bar{B}^0_s \to J/\psi\gamma) = 2.11\times10^{-7}$,
$\mathcal{B}(\bar{B}^0_d \to J/\psi\gamma) =7.65\times10^{-9}$.
Future measurements of these decays would be useful for testing
the factorization frameworks.  The smallness of these decays in
the SM make them sensitive probes of New Physics. As an example,
we will consider the possible admixture of (V+A) charge current to
the standard (V-A) current. This admixture will give  significant
contributions to the decays.

\end{abstract}

\vspace{2.0cm} \noindent {\bf PACS Numbers 13.25.Hw 12.15.-y
12.38.Bx 12.60.-i}

\newpage
\section{Introduction }
Decays of B mesons to final states containing charmonium
constitute a very sensitive laboratory for the study of
electro-weak interactions, as well as the dynamics of strong
interactions. The semi-inclusive $J/\psi$ productions in B decays
had ever risen tough  challenges for understanding its large
rate\cite{cleo1,babar1} which require large contribution beyond
color singlet model for charmonium production at $m_b$
scale\cite{braaten, beneke1, beneke2, palmer,ko}. Currently its
large production rate could be understood in the framework of
Non-Relativistic QCD(NRQCD) effective field theory\cite{nrqcd}.
However, the momentum spectrum of $J/\psi$, especially the excess
of slow $J/\psi$ meson, is still hard for theoretical explanation,
which may reveal interesting  phenomena of the possible intrinsic
charm component of the B\cite{hou}, the decay $B\to J/\psi$ baryon
anti-baryon\cite{brodsky} and the production of $s{\bar d}g$
hybrid\cite{yangeilam2}. The exclusive decays $B\to J/\psi
K^{(*)}$ have also attracted very extensive theoretical and
experimental studies, which involve more complicated strong
dynamics. The recent studies\cite{cheng, chay} have shown that it
is hard to account  for its large rates and polarizations
theoretically.

In this paper, we present a study of the radiative annihilation
decays $\bar{B}^0_{s,d}\to J/\psi\gamma$, which is much rarer than
$B\to J/\psi K^{(*)}$, however, involve simpler hadronic dynamics.
In naive factorization approach, these decays involve the form
factors which are similar to those of the radiative leptonic
decays\cite{tmyan, sach, pirjol, kou}. It is shown that the form
factors could be described simply in terms of a convolution of the
B meson distribution function with a perturbative
kernel\cite{sach}. Beyond naive factorization approach,
non-factorization contributions only arise from one loop vertex
QCD corrections which can be calculate properly  in the framework
of QCD factorization\cite{bbns1}. We find ${\cal B}(\bar{B}^0_s
\to J/\psi \gamma)=1.40\times10^{-6}$ and ${\cal B}(\bar{B}^0_d
\to J/\psi \gamma)=5.29\times10^{-8}$ in naive factorization
approach. With QCD factorization approach, these branching ratios
are reduced  to be ${\cal B}(\bar{B}^0_s \to J/\psi
\gamma)=2.11\times10^{-7}$ and ${\cal B}(\bar{B}^0_d \to J/\psi
\gamma)=7.65\times10^{-9}$, because the effective coefficient
$a_2$ gets smaller than it in naive factorization. Such  effects
have also been found in two body B non-leptonic
decays\cite{bbns1,bbnspk,yang,du} where the vertex-type one loop
QCD corrections make  $|a_2 |$ much smaller than it naive one,
i.e., $a_2 =C_2 +C_{1}/N_{c}$. Compared with two body non-leptonic
B decays,  the troublesome hard spectator scattering contribution
is absent and only the well defined vertex type nonfactorizable
corrections are encountered.  To this extend, the decays
$\bar{B}^0_{s,d} \to J/\psi\gamma$ could be used to test
factorization schemes. On the other hand, these decays could serve
as probes for new physics activities as low energy scale. As an
example, we take these decays as probes of the chirality of weak
currents induced the decays $b\to c \bar{c} s(d)$.

The paper is organized as follows. In Sec.2, we present our study
of $\bar{B}^0_{s,d}\to J/\psi \gamma$ in the standard model(SM).
In Sec.3, we will calculate the effect of the possible admixture
of $(V+A)$ current $g_{R}(\bar{q}_1q_2)_{V+A}$ to the standard
$(V-A)$ current $g_{L}$$(\bar{q}_1q_2)_{V-A}$ in the decays
$\bar{B}^0_{s,d} \to J/\psi$ $\gamma$. This admixture could lead
to enhancement of the decays branching ratios. The effects of the
small value of $g_{R}/g_{L}$ could show up signals of the activity
of New Physics.

\section{$\bar{B}^0_{s,d} \to J/\psi$ $\gamma$ in the SM}

We start our study from the effective Hamiltonian relevant to
$\bar{B}^0_q \to J/\psi$ $\gamma$ decays in the SM \cite{buras}
\begin{eqnarray}
 \mathcal{H}_{eff}=\frac{G_{F}}{\sqrt{2}}\biggl\{V_{cb} V^{*}_{cq}
 \Big[C_{1}(\mu)\mathcal{O}^{c}_{1}(\mu)+C_{2}(\mu)\mathcal{O}^{c}_{2}(\mu)\Big]
 -V_{tb}
 V^{*}_{tq}\sum_{i=3}^{10}C_{i}(\mu)\mathcal{O}_{i}(\mu)\biggr\}.
\end{eqnarray}
Where q=s,d and $C_{i}$ (i=1,$\cdots$,10) are the effective Wilson
coefficients at next-to-leading order evaluated at the
renormalization scale $\mu$. The effective operators can be
expressed explicitly as follows\cite{buras}
\begin{eqnarray}
\mathcal{O}^{c}_{1}&=&(\bar{c}_{\alpha}b_{\beta})_{V-A}
\otimes(\bar{q}_{\beta}c_{\alpha})_{V-A},\hspace{1cm}
\mathcal{O}^{c}_{2}  \hspace{0.35cm} =\hspace{0.35cm}
(\bar{c}_{\alpha}b_{\alpha})_{V-A}
\otimes(\bar{q}_{\beta}c_{\beta})_{V-A},\nonumber\\
\mathcal{O}_{3}&=&(\bar{q}_{\alpha}b_{\alpha})_{V-A}
\otimes(\bar{c}_{\beta}c_{\beta})_{V-A},\hspace{1cm}
\mathcal{O}_{4}\hspace{0.35cm}=\hspace{0.35cm}(\bar{q}_{\alpha}b_{\beta})_{V-A}
\otimes(\bar{c}_{\beta}c_{\alpha})_{V-A},\nonumber\\
\mathcal{O}_{5}&=&(\bar{q}_{\alpha}b_{\alpha})_{V-A}
\otimes(\bar{c}_{\beta}c_{\beta})_{V+A},\hspace{1cm}
\mathcal{O}_{6}\hspace{0.35cm}=\hspace{0.35cm}(\bar{q}_{\alpha}b_{\beta})_{V-A}
\otimes(\bar{c}_{\beta}c_{\alpha})_{V+A},\\
\mathcal{O}_{7}&=&\frac{3}{2}(\bar{q}_{\alpha}b_{\alpha})_{V-A}
\otimes e_{c}(\bar{c}_{\beta}c_{\beta})_{V+A},\hspace{0.37cm}
\mathcal{O}_{8}\hspace{0.35cm}=\hspace{0.35cm}\frac{3}{2}(\bar{q}_{\alpha}b_{\beta})_{V-A}
\otimes e_{c}(\bar{c}_{\beta}c_{\alpha})_{V+A},\nonumber\\
\mathcal{O}_{9}&=&\frac{3}{2}(\bar{q}_{\alpha}b_{\alpha})_{V-A}
\otimes e_{c}(\bar{c}_{\beta}c_{\beta})_{V-A},\hspace{0.37cm}
\mathcal{O}_{10}\hspace{0.21cm}=\hspace{0.35cm}\frac{3}{2}(\bar{q}_{\alpha}b_{\beta})_{V-A}
\otimes e_{c}(\bar{c}_{\beta}c_{\alpha})_{V-A}\nonumber.
\end{eqnarray}
Where $\alpha$ and $\beta$ are the SU(3) color indices.

\begin{figure}[htbp]
\begin{center}
\includegraphics[scale=1.2]{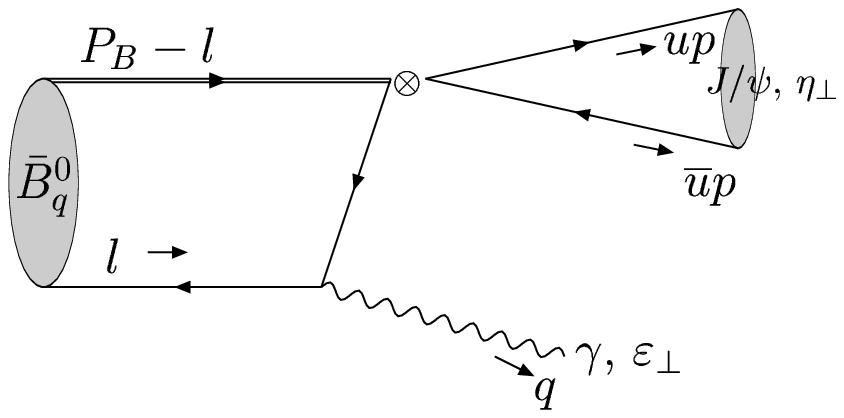}
\caption{ The Feynman diagram for the leading contribution to
$\bar{B}^0_q \to J/\psi$ $\gamma$ decays. The photon radiated from
other quarks are suppressed by power of
$\mathcal{O}(\Lambda_{QCD}/m_{b})$.}
\end{center}
\end{figure}

Under naive factorization, the $\bar{B}^0_{q} \to J/\psi$ $\gamma$
decays are represented by Fig.1. The dominant mechanism is the
radiation of the photon from the light quark in the B meson.
Generally the amplitude is suppressed by one order of
$\Lambda_{QCD}/m_{b}$ because $J/\psi$ meson must be transversely
polarized and B meson is heavy. Radiation of the photon from the
remaining three quark lines is further suppressed additionally  by
power of  $(\Lambda_{QCD}/m_{b})$, which will be neglected in this
paper.

In the heavy quark limit, the decay amplitude of $\bar{B}^0_q \to
J/\psi$ $\gamma$ to leading order  is
\begin{eqnarray}
A( \bar{B}^0_q \to J/\psi \gamma)=\frac{G_{F}}{\sqrt{2}}V_{cb}
V^{*}_{cq}\bar{a}_q
 \sqrt{4\pi\alpha_{e}}f_{J/\psi}M_{J/\psi}F_{V}\nonumber\hspace{4cm}\\
\times\biggl\{-\epsilon_{\mu\nu\rho\sigma}\eta^{\mu}_{\perp}\varepsilon^{\nu}_{\perp}
v^{\rho}q^{\sigma}+i
\Big[(\varepsilon_{\perp}{\cdot}\eta_{\perp})(v{\cdot}q)
-(\eta_{\perp}{\cdot}q)(\varepsilon_{\perp}{\cdot}v)\Big]\biggr\},
\end{eqnarray}
where the approximation $V_{tb}V^*_{tq}\approx-V_{cb}V^*_{cq}$ has
been made, the parameter $\bar{a}_q$ is defined by
\begin{eqnarray}
\bar{a}_q=a_2+a_3+a_5+a_7+a_9,
\end{eqnarray}
with $a_{2i}=C_{2i}+\frac{1}{N_c}C_{2i-1}$ and
$a_{2i-1}=C_{2i-1}+\frac{1}{N_c}C_{2i}$.  $\varepsilon_\bot$ and
$\eta_\bot$ are transverse polarization vectors of photon and
$J/\psi$ meson respectively. The form factor $F_{V}$ is defined by
\cite{tmyan, sach, pirjol, kou}
\begin{eqnarray}
\langle
 \gamma(\varepsilon_{\perp},q)|(\bar{q}b)_{V-A}|\bar{B}^{0}_{q}\rangle =
 \sqrt{4\pi\alpha_{e}}
   \Big[-F_{V}\epsilon_{\mu\nu\rho\sigma}\varepsilon^{\nu}_{\perp}
v^{\rho}q^{\sigma}+iF_{A}(\varepsilon_{\perp\mu}v{\cdot}q
-q_{\mu}\varepsilon_{\perp}{\cdot}v)\Big].
\end{eqnarray}
At leading order of $\mathcal{O}(1/m_{b})$, the two form factors
are given by
\begin{eqnarray}
F_{A}=F_{V}=\frac{Q_{q}f_{B}M_{B}}{2\sqrt{2}E_{\gamma}\lambda_{B}},
\end{eqnarray}
and $\lambda_{B}$ is the first inverse moment of the B-meson's
distribution amplitude
\begin{eqnarray}
\frac{1}{\lambda_{B}}=\int^\infty_0 d l_{+}
\frac{\phi^{B}_{1}(l_{+})}{l_{+}}.
\end{eqnarray}
Now we can write down the helicity amplitude
\begin{eqnarray}
\mathcal{M}_{++}&=&-\frac{G_{F}}{\sqrt{2}}V_{cb}
V^{*}_{cq}\bar{a}_{q}\sqrt{4\pi\alpha_{e}}
F_{V}f_{J/\psi}M_{J/\psi}(2iE_{\gamma}),\nonumber \\
\mathcal{M}_{--} &=& 0. \label{eq1}
\end{eqnarray}
From the helicity amplitude in Eq.\ref{eq1}, the  branching ratio
reads
\begin{eqnarray}
\mathcal{B}(\bar{B}^0_{q} \to J/\psi\gamma) =
\frac{\tau_{B_q}|P_{c}|}{8\pi M^{2}_{B}}\Big(
|\mathcal{M}_{--}|^{2}+|\mathcal{M}_{++}|^{2}\Big),
\end{eqnarray}
here $P_{c}$ is the c.m. momentum and $\tau_{B_q}$ is the lifetime
of $\bar{B}^0_q$ meson.

For numerical analysis, we use the following input
parameters\cite{PDG}:
\begin{center}
\begin{tabular}{llll}
$M_{B_s} = 5.370 Gev,$&$\tau_{B_{s}} = 1.461 ps,$&$m_{b} = 4.8 Gev,$&$V_{cb} = 0.0412,$\\
$M_{B_d} = 5.279 Gev,$&$\tau_{B_{d}} = 1.542 ps,$&$m_{c} = 1.47 Gev,$&$V_{cd} = 0.224,$\\
$M_{J/\psi} = 3.097 Gev,$&$\lambda_{B} = 0.35 Gev \cite{sach,bbns1},$&$N_{c} = 3,$&$V_{cs} = 0.996,$\\
\end{tabular}
\end{center}
the decay constants: $f_{B_s}=210 Mev,$ $ f_{B_d}=180 Mev,$ $
f_{J/\psi}=405 Mev,$ and the Wilson coefficients at $\mu=m_b$
scale: $C_{1}=1.082,
C_{2}=-0.185,C_3=0.014,C_4=-0.035,C_5=0.009,C_6=-0.041,C_7=-\frac{0.002}{137},
C_8=\frac{0.054}{137},C_9=-\frac{1.292}{137},C_{10}=-\frac{0.263}{137}$\cite{buras}.

In naive factorization, we get the branching ratios
\begin{eqnarray}
\mathcal{B}(\bar{B}^0_s \to J/\psi\gamma) &=&
1.40\times10^{-6},\\
\mathcal{B}(\bar{B}^0_d \to J/\psi\gamma) &=& 5.29\times10^{-8}.
\end{eqnarray}

In the above calculations, non-factorizable contributions are
neglected. However, the non-factorizable contributions may be
important. These radiative corrections at order $\alpha_s$ can be
obtained by calculating the amplitudes in Fig.2. The QCD
factorization approach advocated recently in \cite{bbns1} allows
us to compute the non-factorizable corrections in the heavy quark
limit.

In our calculation, we take the momentum of the B meson
$P^{\mu}_{B}=M_{B}v^{\mu}$ and photon flying along
$n_{-}=(1,0,0,-1)$ direction, where the four-velocity
$v=(1,0,0,0)$ satisfies $v^{2}=1$. In the heavy quark limit, the
B-meson's light-cone projection operator  can be written
as\cite{bbnspk,grozin}
\begin{eqnarray}
M^{B}_{\alpha\beta} =
\frac{i}{4N_{c}}f_{B}M_{B}\biggl\{(1+\rlap/v)\gamma_{5}
\Big[\Phi^B_{1}(\rho)+\rlap/n_{-}
\Phi^B_{2}(\rho)\Big]\biggr\}_{\alpha\beta},
\end{eqnarray}
where $\rho$ is the momentum fraction carried by the spectator
quark of the B meson and the normalization conditions are
\begin{eqnarray}
\int^{1}_{0}d\rho\Phi^{B}_{1}(\rho)=1,\hspace{1.5cm}
\int^{1}_{0}d\rho\Phi^{B}_{2}(\rho)=0.
\end{eqnarray}
For $J/\psi$ meson,  we take
\begin{eqnarray}
M^{J/\psi}_{\perp\rho\sigma}=-\frac{f_{J/\psi}}{4N_{c}}
\Big[\rlap/\eta_\perp(\rlap/P_{J/\psi}+M_{J/\psi})
\Big]_{\rho\sigma}\Phi^{J/\psi}(u).
\end{eqnarray}
 Because the charm quark is
heavy, the wave function $\Phi^{J/\psi}(u)$ is symmetric function
under $u \to 1-u$ and should be sharply peaked around $u=1/2$
\cite{yangeilam2}.

\begin{figure}[htbp]
\begin{center}
\includegraphics[scale=0.65]{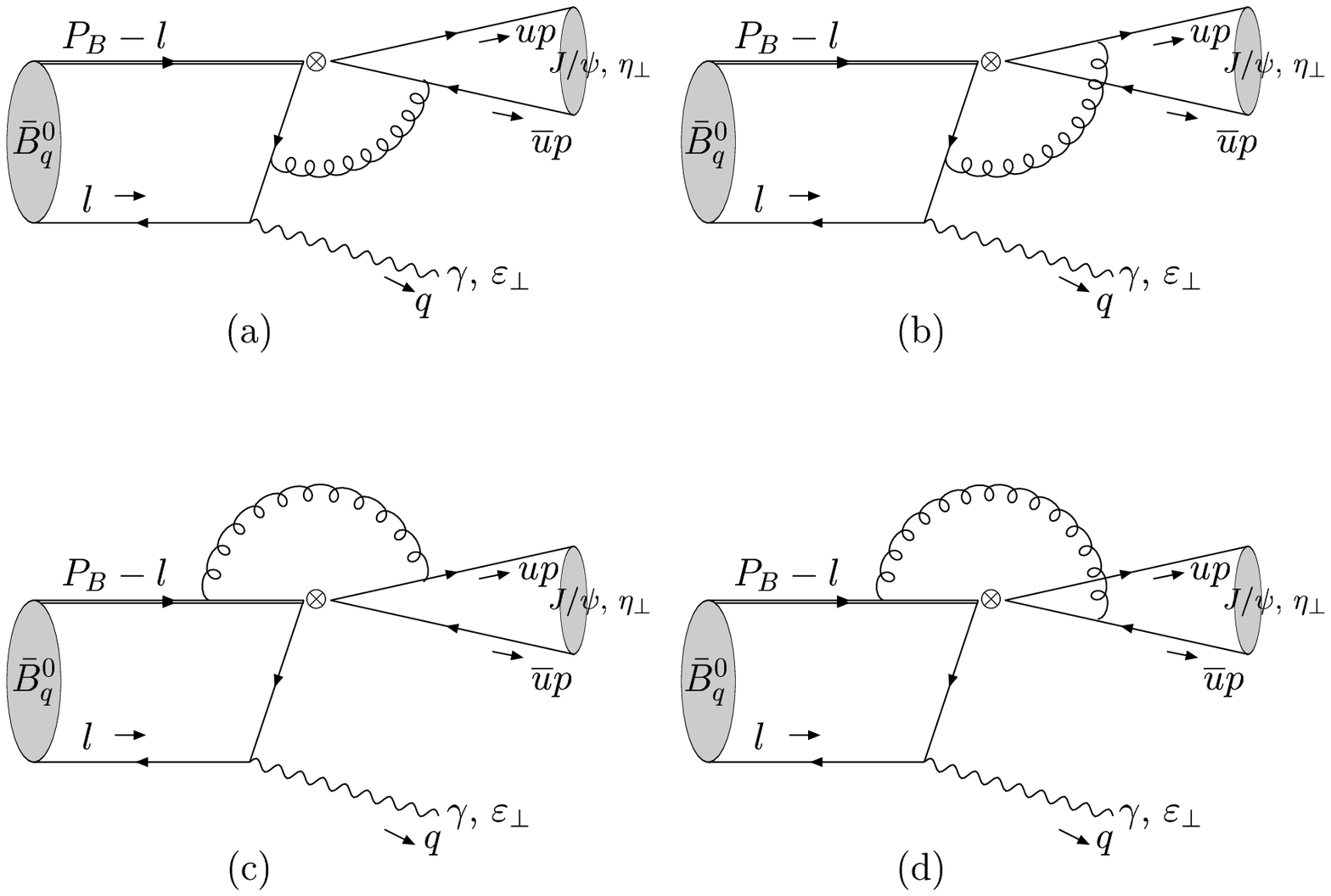}
\caption{Non-factorizable contribution at order
$\alpha_{s}$."$\otimes$" denote the insertions of color-octet
operators
$\mathcal{O}^c_1,\mathcal{O}_4,\mathcal{O}_6,\mathcal{O}_8,\mathcal{O}_{10}$
in $\bar{B}^0_q \to J/\psi$ $\gamma$. Other diagrams with the
photon radiating from remain three quark lines is suppressed and
neglected.}
\end{center}
\end{figure}
The calculation of the non-factorizable contributions depicted in
Fig.2. is straightforward. Including the contributions, the
amplitudes for $\bar{B}^0_q \to J/\psi$ $\gamma$ will be
\begin{eqnarray}
A( \bar{B}^0_q \to J/\psi \gamma)=\frac{G_{F}}{\sqrt{2}}V_{cb}
V^{*}_{cq}\bar{a}'_{q}\sqrt{4\pi\alpha_{e}}f_{J/\psi}M_{J/\psi}
F_{V}\hspace{2cm}\nonumber\\
\times\biggl\{-\epsilon_{\mu\nu\rho\sigma}
\eta^{\mu}_{\perp}\varepsilon^{\nu}_{\perp}
v^{\rho}q^{\sigma}+i\Big[(\varepsilon_{\perp}{\cdot}\eta_{\perp})(v{\cdot}q)
-(\eta_{\perp}{\cdot}q)(\varepsilon_{\perp}{\cdot}v)\Big]\biggr\},
\end{eqnarray}
and
\begin{eqnarray}
\bar{a}'_{q}=a'_2+a'_3+a'_5+a'_7+a'_9.
\end{eqnarray}

The $\mathcal{O}(\alpha_{s})$ corrections are summarized in
$a'_{i}$ which are calculated to be
\begin{eqnarray}
a'_{2}&=&C_{2}+
\frac{C_{1}}{N_{c}}+\frac{\alpha_{s}}{4\pi}\frac{C_F}{N_{c}}C_{1}F,\nonumber\\
a'_{3}&=&C_{3}+
\frac{C_{4}}{N_{c}}+\frac{\alpha_{s}}{4\pi}\frac{C_F}{N_{c}}C_{4}F,\nonumber\\
a'_{5}&=&C_{5}+
\frac{C_{6}}{N_{c}}-\frac{\alpha_{s}}{4\pi}\frac{C_F}{N_{c}}C_{6}F,\\
a'_{7}&=&C_{7}+
\frac{C_{8}}{N_{c}}-\frac{\alpha_{s}}{4\pi}\frac{C_F}{N_{c}}C_{8}F,\nonumber\\
a'_{9}&=&C_{9}+
\frac{C_{10}}{N_{c}}+\frac{\alpha_{s}}{4\pi}\frac{C_F}{N_{c}}C_{10}F.\nonumber
\end{eqnarray}
where $C_{F}=(N_{c}^2-1)/(2N_{c})$, the $\alpha_s$ terms are the
non-factorizable contributions which comes from one gluon exchange
between the two currents of color-octet  operators
$\mathcal{O}^c_{1},\mathcal{O}_4,\mathcal{O}_6,\mathcal{O}_8,\mathcal{O}_{10}$,
as shown by Fig.2. We get
\begin{eqnarray}
F&=&-16-12ln\frac{\mu}{M_B}-i\pi-\int^{1}_{0}du\Phi^{J/\psi}(u)
\left\{-\left( 1+\frac{uz}{1-(1-u)z}\right) ln(1-z)
\right.  \nonumber\\
&&+\frac{4u-5}{1-u}ln(u)
+\left(\frac{3}{1-uz}+\frac{1}{1-(1-u)z}
  \right)u z ln(uz)
  +\frac{i\pi uz}{1-(1-u)z}\nonumber
\\
&&+2tz\left[
\frac{ln(u)}{z(1-u)}+\frac{ln(1-z)-i\pi}{1-(1-u)z}+\left(\frac{1}
{1-uz}-\frac{1}{1-(1-u)z}\right)ln(uz)
      \right] \nonumber \\
      &&
      +2\left.\left( \frac{t}{u}-1\right)
      \left[ Li_{2}\left(\frac{u-1}{u}\right)
-Li_{2}\left(\frac{1-uz}{uz}\right)
+Li_{2}\left(\frac{1-(1-u)z}{uz}\right)+\gamma_{E}
       \right]  \right\}.
\end{eqnarray}
Where $z=\frac{M_{J/\psi}^{2}}{M_{B}^{2}}$ and
$t=\frac{m_{c}}{M_{J/\psi}}$. We have neglected the difference
between $m_{b}$ and $M_{B}$ which is a sub-leading effect in the
heavy quark limit.

In the calculation, the $\overline{MS}$ renormalization scheme is
used. We neglect the small effect of box diagrams and also
neglected $l^{2}_{+}$ terms entered in the loop calculation which
are the higher twist effect. Therefore, the integral involved
$\Phi^{B}_{2}(\rho)$ absents and remain integrals could be related
to the form factor $F_{V}$.

To order of $\alpha_{s}$ corrections, the helicity amplitudes are
\begin{eqnarray}
\mathcal{M}_{++}&=&-\frac{G_{F}}{\sqrt{2}}V_{cb}
V^{*}_{cq}\bar{a}'_{q}\sqrt{4\pi\alpha_{e}}
F_{V}f_{J/\psi}M_{J/\psi}(2iE_{\gamma}),\\
\mathcal{M}_{--} &=& 0.
\end{eqnarray}

For $\Phi^{J/\psi}(u)=6u(1-u)$, we  obtain
\begin{eqnarray}
{\cal B}(\bar{B}^0_s \to J/\psi\gamma) &=& 2.11\times10^{-7},\\
{\cal B}(\bar{B}^0_d \to J/\psi\gamma) &=& 7.65\times10^{-9}.
\end{eqnarray}
Because the shape of the $J/\psi$ wave function is unknown, it is
worth considering other possibilities besides the asymptotic form.
The delta-function form $\Phi^{J/\psi}(u)=\delta(u-\frac{1}{2})$
as used in \cite{yangbc}, appeals to the naive expectation of the
wave function in the non-relativistic limit. Using this wave
function, the results are
\begin{eqnarray}
\mathcal{B}(\bar{B}^0_s \to J/\psi\gamma) &=& 9.09\times10^{-8},\\
\mathcal{B}(\bar{B}^0_d \to J/\psi\gamma) &=& 3.22\times10^{-9}.
\end{eqnarray}

The $\bar{B}^0_s \to J/\psi\gamma$ decays may be measured at
Tevatron and LHC in the future. The decays $\bar{B}^0_d \to
J/\psi\gamma$ could be studied  in the planning super B factories
at KEK and SLAC succeeded to Belle and BaBar. Potentially  they
could be enhanced by New Physics and the enhancement might be
measured at those facilities.

\section{ An admixture of $(V+A)$ current in the
$\bar{B}^0_{s,d} \to J/\psi\gamma$}

B decays are known to be governed by  weak couplings and small
mixing matrix elements. These decays are therefore very sensitive
to new kinds of interactions and in particular to right-handed
couplings \cite{M.Gronau}.
 It is conventionally assumed
that the B-decays proceed via the pure $(V-A)$ current in the SM.
However, it turns out to be surprisingly difficult to exclude the
possibility  that the dominant B decays occur via a $(V+A)$
coupling\cite{9209278}. The (V+A) coupling has been studied in
Ref. \cite{9705261,9704278,9704287}. In the annihilation B-decays
$\bar{B} ^0_{s,d} \to J/\psi$ $\gamma$, there is the possible
admixture of $(V+A)$ charged current $g_R(\bar{q}_1q_2)_{V+A}$ to
the standard $(V-A)$ current $g_L(\bar{q}_1q_2)_{V-A}$. The small
value of $g_R/g_L$ is not ruled out so far and can be sought for
as one possible sign of New Physics. In what following, we will
examine the effect of possible admixture.

Assuming  the admixture of $(V+A)$ charge currents ($b \to c$) and
($c \to q$) to the SM $(V-A)$ currents, the effective
four-fermion interaction operators for $\bar{B}^0_q \to J/\psi$
$\gamma$ here can be written as
\begin{eqnarray}
\mathcal{O}^{c}_{1}&=&\Big[(\bar{c}_{\alpha}b_{\beta})_{V-A}
+\xi(\bar{c}_{\alpha}b_{\beta})_{V+A}\Big]\otimes
\Big[(\bar{q}_{\beta}c_{\alpha})_{V-A}
+\xi'(\bar{q}_{\beta}c_{\alpha})_{V+A}\Big],\nonumber\\
\mathcal{O}^{c}_{2}&=&\Big[(\bar{c}_{\alpha}b_{\alpha})_{V-A}
+\xi(\bar{c}_{\alpha}b_{\alpha})_{V+A}\Big]\otimes
\Big[(\bar{q}_{\beta}c_{\beta})_{V-A}+\xi'(\bar{q}_{\beta}c_{\beta})_{V+A}\Big],
\end{eqnarray}
with $\xi=g_R/g_L$ for current ($b \to c$), $\xi'=g'_R/g'_L$ for
current ($c \to q$). Here we use the approximation $\bar{a}_q=a_2$
and $\bar{a}'_q=a'_2$.

The  amplitude for $\bar{B}^0_q \to J/\psi\gamma$ is
\begin{eqnarray}
A( \bar{B}^0_q \to J/\psi \gamma)&=&\frac{G_{F}}{\sqrt{2}}V_{cb}
V^{*}_{cq}\sqrt{4\pi\alpha_{e}}f_{J/\psi} M_{J/\psi}F_{V}
\nonumber \\
&\times&\biggl\{(1+\xi')a_{2}
\Big[-(1+\xi)\epsilon_{\mu\nu\rho\sigma}\eta^{\mu}_{\perp}
\varepsilon^{\nu}_{\perp}v^{\rho}q^{\sigma}
+i(1-\xi)(\varepsilon_{\perp}{\cdot}\eta_{\perp})(v{\cdot}q)
\Big] \biggr.\\
&+&\biggl. (1-\xi') \frac{\alpha_s}{4\pi}\frac{C_F}{N_C}C_1F
\Big[(\xi-1)\epsilon_{\mu\nu\rho\sigma}\eta^{\mu}_{\perp}
\varepsilon^{\nu}_{\perp}v^{\rho}q^{\sigma}
+i(1+\xi)(\varepsilon_{\perp}{\cdot}\eta_{\perp})(v{\cdot}q) \Big]
\biggr\}. \nonumber
\end{eqnarray}
The branching ratios can be read
\begin{eqnarray}
\mathcal{B}(\bar{B}^0_{q} \to J/\psi\gamma)=
\frac{\tau_{B_q}|P_{c}|}{8\pi M^{2}_{B}}
\frac{G^2_{F}}{2}\left|V_{cb} V^{*}_{cq}\right|^2
4\pi\alpha_{e}f^2_{J/\psi}M^2_{J/\psi}F^2_{V}4E^2_{\gamma}
\hspace{3.5cm}
\nonumber\\
\times\biggl\{ \Big| (1+\xi')a_{2}+(1-\xi')\frac{\alpha_{s}}
{4\pi}\frac{C_F}{N_{C}}C_{1}F\Big|^{2} + \xi^2 \Big|
(1+\xi')a_{2}-(1-\xi')\frac{\alpha_{s}}
{4\pi}\frac{C_F}{N_{C}}C_{1}F\Big|^{2} \biggr\}.
\end{eqnarray}
Where the factor $\xi^2$ stems from $\xi(\bar{c}b)_{V+A}$. Because
$\xi$ is a small constant, in what fellows we will neglect
$\xi^{2}$.

\begin{figure}[htbp]
\begin{center}
\begin{tabular}{cc}
\includegraphics[scale=0.9]{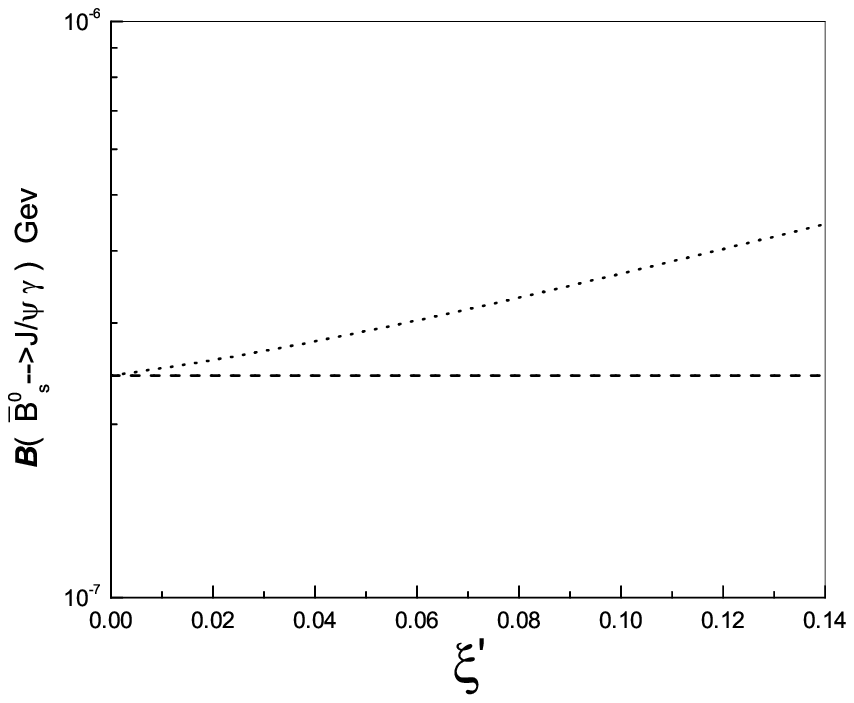}&
\includegraphics[scale=0.9]{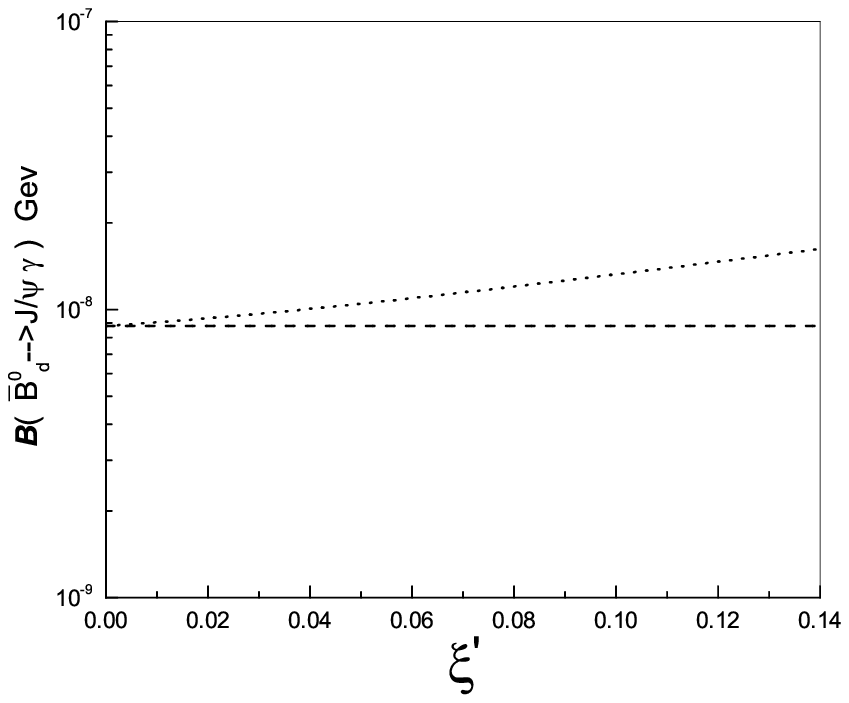}
\end{tabular}
\end{center}
\caption{$\mathcal{B}(\bar{B}^0_{s,d} \to J/\psi$ $\gamma)$ as a
function of $\xi'=g'_R/g'_L$. The dash lines are the SM
predictions and the dot lines are the results of including a small
admixture of (V+A) quark current.}
\end{figure}

Using the $\Phi^{J/\psi}(u)=6u(1-u)$, branching ratios are
\begin{eqnarray}
\mathcal{B}(\bar{B}^0_s \to J/\psi\gamma) &=&
(0.46+1.27\xi'+10.61\xi'^{2})\times5.23\times10^{-7},\\
\mathcal{B}(\bar{B}^0_d \to J/\psi\gamma) &=&
(0.44+1.19\xi'+10.71\xi'^{2})\times1.98\times10^{-8}.
\end{eqnarray}
We can see that the branching ratios of these decays are very
sensitive to the possible presence of the admixture of right hand
current. Comparing with the SM predictions, these decays are
enhanced. The numerical results are displayed as an illustration
in Fig.3.

\section{Conclusion}

 We have studied the radiative annihilation
 decays $\bar{B} ^0_{s,d} \to J/\psi$ $\gamma$ within the framework
of QCD factorization approach. Physically, the factorization
method is  applicable because the transverse size of $J/\psi$ is
small in the heavy quark limit. We have shown that the
non-factorizable radiative corrections
 at order $\alpha_{s}$ change the magnitude significantly
 compared to the leading-order result corresponding to the naive
 factorization. In naive factorization, we find
 $\mathcal{B}(\bar{B}^0_s \to J/\psi\gamma)=1.40\times 10^{-6}$
 and $\mathcal{B}(\bar{B}^0_d \to J/\psi\gamma)=5.29\times
 10^{-8}$ which are much larger than
$\mathcal{B}(\bar{B}^0_s \to J/\psi\gamma)=2.11\times 10^{-7}$
 and $\mathcal{B}(\bar{B}^0_d \to J/\psi\gamma)=7.65\times
 10^{-9}$ in the framework of QCD factorization at order of
 $\alpha_s$. It is interesting to note that these decays involve simpler
 hadronic dynamics than two-body B non-leptonic decays.
  Experimental measurements  would be very useful for
  understanding the mechanics of $J/\psi$ productions and testing
  the factorization frameworks. Simultaneously these decays also
  are the background for the interesting decays
  $\bar{B}^0_{s,d}\to \mu^+ \mu^- \gamma$. On the other hand,
  these decays may be sensitive to New Physics. As an
  illustration, we have investigated the effects of the admixture
  of right-hand currents.
 We find that these decays are sensitive to admixture of the
right-handed ($c \to s,d$) current and the effect of the admixture
of  right-handed ($b \to c$) current is negligible small.
Experimentally   these decays could be studied at CERN LHC and the
planning super high luminosity B factories at KEK and SLAC.
\vspace{1cm}\\
 \noindent{\bf {\Large Acknowledgments }}

 Y.D is supported by the Henan Provincial Science
Foundation for Prominent Young Scientists under the contract
0312001700. This work is supported in part by National Science
Foundation of China under the contracts 19805015 and 1001750.

{}

\end{document}